\documentclass[12pt,preprint]{aastex}
\usepackage{emulateapj5,apjfonts}
\usepackage{graphics}
\usepackage{amssymb}
\usepackage{onecolfloat}
\lefthead{Heitmann, Luki\'c, Habib, Ricker}
\righthead{Capturing Halos at High Redshifts}

\begin{document}


\def\head{
  \vbox to 0pt{\vss
                    \hbox to 0pt{\hskip 440pt\rm LA-UR-05-9198\hss}
                   \vskip 25pt}

\title{Capturing Halos at High Redshifts}
\author{Katrin~Heitmann\altaffilmark{1},
        Zarija~Luki\'c\altaffilmark{2}, 
        Salman~Habib\altaffilmark{3}, and         
        Paul~M.~Ricker\altaffilmark{2,4}}

\affil{$^1$ ISR-1, ISR Division, The University of California, Los
Alamos National Laboratory, Los Alamos, NM 87545}
\affil{$^2$ Dept.\ of Astronomy, University of Illinois, Urbana, IL 61801}
\affil{$^3$ T-8, Theoretical Division, The University of California, Los
Alamos National Laboratory, Los Alamos, NM 87545}
\affil{$^4$ National Center for Supercomputing Applications, Urbana, IL 61801}

\date{today}

\begin{abstract}

We study the evolution of the mass function of dark matter halos in
the concordance $\Lambda$CDM model at high redshift. We employ
overlapping (multiple-realization) numerical simulations to cover a
wide range of halo masses, $10^7-10^{15}h^{-1}M_\odot$, with redshift
coverage beginning at $z=20$. The Press-Schechter mass function is
significantly discrepant from the simulation results at high
redshifts. Of the more recently proposed mass functions, our results
are in best agreement with Warren et al. (2005). The statistics of the
simulations -- along with good control over systematics -- allow for
fits accurate to the level of $20\%$ at all redshifts. We provide a
concise discussion of various issues in defining and computing the
halo mass function, and how these are addressed in our simulations.

\end{abstract}

\keywords{methods: N-body simulations ---
          cosmology: halo mass function}}

\twocolumn[\head]
\section{Introduction}

Dark matter halos occupy a central place in the paradigm of
gravitationally-driven structure formation arising from the nonlinear
evolution of primordial Gaussian density fluctuations. Gas
condensation, resultant star formation, and eventual galaxy formation
occurs within halos. Consequently, the halo profile and mass function
are central ingredients in phenomenological models of nonlinear
clustering of galaxies. The distribution of halo masses -- the halo
mass function -- and its time evolution, are also sensitive probes of
cosmology.

The halo mass function at the high-mass end (cluster mass scales) is
exponentially sensitive to the amplitude of the initial density
perturbations, the mean matter density parameter, $\Omega_{m}$, and to
the dark energy controlled late-time evolution of the density
field. The last feature, particularly at low redshifts, $z<2$, allows
cluster observations to constrain the dark energy content,
$\Omega_{\Lambda}$, and the equation of state parameter,
$w$~\cite{Holder01}.

The halo mass function is also of considerable interest at high
redshift, relating to questions such as predictions for quasar
abundance and formation sites~\cite{HaimanLoeb01}, the formation
history of collapsed baryonic halos, and the reionization history of
the Universe~\cite{Furlanetto05}. Recent results from the Wilkinson
Microwave Anisotropy Probe (WMAP) (Kogut et al. 2003; Spergel et
al. 2003) indicate that reionization could have begun at redshifts as
high as $z\sim 20$. Much of the work on possible reionization
scenarios is based on the simple Press-Schechter (PS) mass
function~(Press \& Schechter~1974, Bond et al. 1991) the use of which
can lead to imprecise predictions for the reionization history.

Simulations play a dual role in characterization of the halo mass
function. If only a few fixed sets of cosmological parameters and a
finite dynamic range are required, simulations can produce valuable
results. In order to investigate a variety of cosmologies and
different scenarios for physical processes, e.g., reionization, it is
nevertheless very convenient, if not necessary, to have accurate
analytic fitting relations. Simulations can be used to validate these
fits over a wide (albeit, discretely sampled) range of parameters.

Various numerical studies of the mass function have been carried out
over different mass and redshift ranges. The closest to the present
work are Reed et al. (2003) and Springel et al. (2005); in comparison
to their results, our halo mass range goes deeper by three orders of
magnitude, with good statistics and control of systematics out to
$z=20$, substantially higher than in these papers. (We review results
from other work below.) Essentially, the earlier results are in very
good agreement with the Sheth-Tormen mass function~\cite{Sheth99}, at
redshifts $z\leq 10$. As we show below, various fitting formulae given
in the literature -- most tuned to simulation results at $z=0$ -- can
differ substantially in their predictions at high redshifts, by as
much as a factor of two. Therefore, it is important to carry out
simulations of sufficient dynamic range and accuracy to test these
predictions.

In order to extract the mass function from simulations, different
questions have to be addressed, such as: How is the mass function
to be defined? When do the first halos form in a
simulation?  When must the simulation be started in order to capture
these halos?  What force and mass resolution is required to capture
halos of a certain mass at a specific redshift? We have derived and
tested certain criteria to ensure that our simulations capture the
halos of interest; details will be given elsewhere~\cite{lukic06}.

The paper is organized as follows. In Section~\ref{massf} we discuss
popular mass function formulae, previous work, and strategies for
determining the halo mass function at high redshifts. The simulations
and mass function results are discussed in
Section~\ref{code}. Criteria for mass and force resolution and initial
redshift needed to span the desired mass and redshift range are given
here. We present our conclusions in Section~\ref{conclusion}.

\section{The Mass Function}
\label{massf}

Over the last three decades different fitting functions for the mass
function have been suggested. The first analytic model for the mass
function was developed by Press \& Schechter (1974). Their theory
considers a spherically overdense region in an otherwise smooth
background density field. The overdensity evolves as a Friedmann
universe with positive curvature. Initially, the overdensity expands,
but at a slower rate than the background universe (thus enhancing the
density contrast), until it reaches the `turnaround' density, after
which it collapses. Although formally this collapse ends with a
singularity, it is assumed that in reality the overdense region will
virialize.  For an Einstein-de Sitter universe, the density of such an
overdense region at the virialization redshift is $\rho\approx 180
\rho_c(z)$.  At this point, the density contrast from the linear
theory of perturbation growth [$\delta(\vec{x},z) = D^{+}(z)
\delta(\vec{x},0)$] is $\delta_c(z) \approx 1.686(1+z)$.  For
$\Omega_m < 1$, $\delta_c(z)$ evolves differently (see Lacey \& Cole
1993), but the dependence on cosmology is weak (see e.g., Jenkins et
al. 2001).  Thus, we adopt $\delta_c = 1.686=\delta_c(0)$.

Following the above reasoning and with the assumption that the initial
density perturbations are given by a Gaussian random field, the PS
mass function is given by:
\begin{equation}
f_{PS}(\sigma) = \sqrt{\frac{2}{\pi}} \frac{\delta_c}{\sigma} 
\exp \left[ - \frac{\delta^2_c}{2\sigma^2} \right],
\end{equation}
where $\sigma$ is the variance of the linear density field, 
$f(\sigma,z)\equiv (M/\rho_b)(dn/d\ln\sigma^{-1})$, and $\rho_b$
is the background density.

Using empirical arguments Sheth \& Tormen (1999, hereafter ST)
proposed an improved fit of the following form:
\begin{equation}
f_{ST}(\sigma) = 0.3222 \sqrt{\frac{2a}{\pi}} \frac{\delta_c}{\sigma} \exp 
\left[ - \frac{a \delta^2_c}{2\sigma^2} \right]
\left[ 1 + \left( \frac{\sigma^2}{a \delta^2_c} \right) ^ p \right],
\end{equation}
with $a=0.707$, and $p=0.3$.  Sheth et al.  (2001) interpreted this
fit theoretically by extending the PS approach to an ellipsoidal
collapse model.  In this model, the collapse of a region depends not
only on its initial overdensity, but also on the surrounding shear
field.  The dependence is chosen to recover the Zel'dovich
approximation \cite{zel70} in the linear regime.  A halo is considered
virialized when the third axis collapses (see also Lee \& Shandarin
1997).

Jenkins et al.  (2001, hereafter Jenkins) combine high resolution
simulations for different cosmologies spanning a mass range of over
three orders of magnitude [$\sim (10^{12} - 10^{15}) h^{-1}M_{\sun}$],
and including several redshifts between $z=5$ and $z=0$.  They find
that the following fitting formula works exceptionally well (within
20\%), independent of the underlying cosmology:
\begin{equation}
f_{Jenkins}(\sigma) = 0.315 \exp 
\left[ -| \ln \sigma^{-1} + 0.61 |^{3.8} \right].
\end{equation}
The above formula is very close to the nominal ST fit.

By performing 16 nested-volume simulations Warren et al.  (2005,
hereafter Warren) obtain significant halo statistics spanning a mass
range of five orders of magnitude~[$\sim (10^{10} - 10^{15})h^{-1}
M_{\sun}$]. Their best fit employs a functional form similar to an
improved version of ST (Sheth \& Tormen 2002):
\begin{equation}
f_{Warren}(\sigma) = 0.7234 \left( \sigma^{-1.625} + 0.2538 \right) 
\exp \left[ -\frac{1.1982}{\sigma^2}\right].
\end{equation}

\begin{figure}
  \includegraphics[width=90mm]{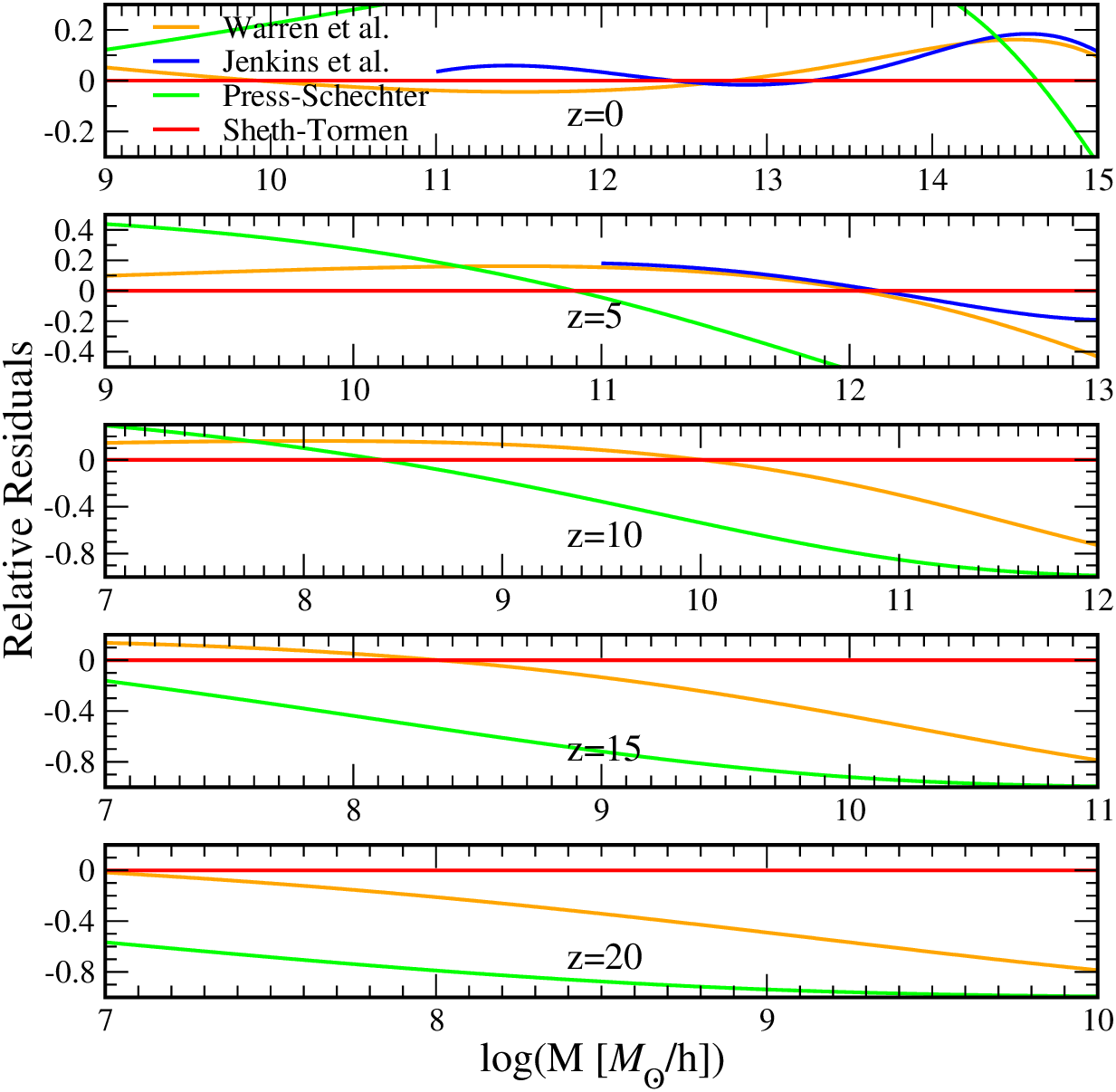}
\caption{Relative residuals of the PS, Jenkins, and Warren mass
function fits with respect to ST for five different redshifts.  Note
that the ranges of the axes are different in the different panels. We
do not show the Jenkins fit below masses of 10$^{11} h^{-1}M_\odot$
since it is not valid in this mass range.}
\label{resids}
\end{figure}

The discrepancy between PS and the more accurate fits is evident in
Figure~\ref{resids} where the redshift evolution of the mass function
is shown. The redshift dependence in the analytic mass functions
enters only through $\sigma(z)=\sigma(0) d(z)$, where $d(z)$ is the
growth factor normalized such that $d(0)=1$. As the functional
dependence on $\sigma$ is different in the different fits, this leads
to substantial variation across the fits as a function of redshift.
For $z=0$ the Warren fit agrees -- especially in the low mass range
below $10^{13} M_\odot$ -- to better than 5\% with the ST fit. At the
high mass end the difference increases up to 20\%. The Jenkins fit
leads to similar results over the considered mass range. Note that at
higher redshifts and intermediate mass ranges around $10^{9}M_\odot$
the disagreement between the Warren and ST fits increases up to 40\%.

The determination of mass functions at high redshifts is a nontrivial
task. High redshift halos have very low masses, placing heavy demands
on the mass and force resolution needed to resolve them. These
requirements can be achieved in two ways.  First, a simulation with a
very large number of particles and high force resolution can be
performed. This is expensive, and only a very limited number of such
simulations can be carried out. Second, since determining the mass
function is simply a question of statistics, many relatively modest
simulations with moderate particle loading can be performed: this is
the strategy we adopt here. As simulations can only be trusted until a
redshift at which the largest mode is close to becoming nonlinear,
multiple overlapping box sizes must be used.

Springel et al. (2005) have recently followed the evolution of
2160$^3$ particles in a $500 h^{-1}$Mpc box.  The high mass and force
resolution allow them to study the mass function reliably out to a
redshift of $z=10$, covering a mass range of roughly
$10^{10}h^{-1}M_\odot$ to $10^{16} h^{-1}M_\odot$.  Examples of {\em
single} small-box simulations include Jang-Condell \& Hernquist
(2001) (1$h^{-1}$Mpc box with $128^3$ particles evolved to $z=10$) and
Cen et al. (2004) ($4h^{-1}$Mpc box, $512^3$ particles, evolved to
$z=6$). Results in both papers are claimed to be consistent with PS but
without detailed quantification.  The simulation of Reed et al. (2003)
is a compromise between the two extremes: a box size of 50$h^{-1}$Mpc
with 432$^3$ particles and a concomitant halo mass range of roughly
10$^{10}h^{-1} M_\odot$ to $10^{14.5}h^{-1}M_\odot$.  Reed et al.
find good agreement (better than 20\%) with the ST fit up to $z\simeq
10$.  For higher redshifts they find that the ST fit overpredicts the
number of halos, at $z=15$ up to 50\%.  At this high redshift,
however, their results become statistics-limited, the mass resolution
being insufficient to resolve the very small halos.

In this paper we analyze a suite of 50 N-body simulations with
varying box sizes between $4h^{-1}$Mpc and $126h^{-1}$Mpc with
multiple realizations of all boxes to study the mass function at
redshifts up to $z=20$ and to cover a large mass range between $10^7
h^{-1}M_\odot$ and $10^{15} h^{-1}M_\odot$ even at high
redshifts. Significantly, at $z=20$, gas in halos with a mass scale
above $\sim 10^7 h^{-1}M_\odot$ can cool via atomic line
cooling~\cite{Tegmark97}.

\section{Simulations and Mass Function Results}
\label{code}
\begin{table*}[t]
\begin{center}
\caption{\label{tabsruns} Summary of the performed runs}
\begin{tabular}{cccccccc}
\hline\hline
 Mesh & Box Size & Resolution & $z_{in}$  & $z_{final}$ &
Particle Mass & Smallest Halo  &\# of Realizations\\
\hline
 1024$^3$ & 126 $h^{-1}$Mpc & 120 $h^{-1}$kpc  & 50  & 0 & $9.94\times 10^{9}~h^{-1}M_\odot$ 
& $3.98 \times 10^{11}~h^{-1}M_\odot$  & 10\\
 1024$^3$ &  64 $h^{-1}$Mpc & 62.5 $h^{-1}$kpc  & 80 & 0 & $1.30\times 10^9~h^{-1}M_\odot$   
& $5.2 \times 10^{10}~h^{-1}M_\odot$ &  5\\
 1024$^3$ &  32 $h^{-1}$Mpc & 31.25 $h^{-1}$kpc & 150 & 5 & $1.63\times 10^8~h^{-1}M_\odot$   
& $6.52 \times 10^{9}~h^{-1}M_\odot$  & 5\\
 1024$^3$ &  16 $h^{-1}$Mpc & 15.63 $h^{-1}$kpc & 200 & 5 &$2.04\times 10^7~h^{-1}M_\odot$  
& $8.16\times 10^{8}~h^{-1}M_\odot$  & 5\\
 1024$^3$ &   8 $h^{-1}$Mpc &  7.81 $h^{-1}$kpc & 250 & 10 & $2.55\times 10^6~h^{-1}M_\odot$ 
& $1.02 \times 10^{8}~h^{-1}M_\odot$ & 20\\
 1024$^3$ &   4 $h^{-1}$Mpc &  3.91 $h^{-1}$kpc & 500 & 10 & $3.19\times 10^5~h^{-1}M_\odot$ 
& $1.27 \times 10^{7}~h^{-1}M_\odot$ & 5\\
\hline\hline

\vspace{-1.5cm}

\tablecomments{Mass and force resolutions of the different runs. The
smallest halos we consider contain 40 particles. All simulations are
run with 256$^3$ particles.}
\end{tabular}
\end{center}
\end{table*}

All simulations in this paper are carried out with the particle-mesh
code MC$^2$ ({\bf M}esh-based {\bf C}osmology {\bf C}ode).  MC$^2$ has
been extensively tested against other cosmological simulation codes
(Heitmann et al. 2005). The chosen values of cosmological parameters are:
\begin{eqnarray}
&&\Omega_{\rm tot}=1.0,~~~\Omega_m=0.253,~~~ 
\Omega_{\rm baryon}=0.048, \nonumber\\
&&\sigma_8=0.9,~~~ H_0=70~{\rm km/s/Mpc},
\end{eqnarray}
as set by the latest cosmic microwave background and large scale
structure observations \cite{MacT05}. The mass transfer functions are
generated with CMBFAST (Seljak \& Zaldarriaga 1996). We summarize the
different runs, including their force and mass resolution in
Table~\ref{tabsruns}.  

We identify halos with the friends-of-friends algorithm (FOF), based
on finding neighbors of particles at a certain distance (see e.g.,
Einasto et al. 1984; Davis et al. 1985). The halo mass is defined
simply by the sum of particles which are members of the halo. (For
connections between different definitions of halo masses, see White
2001.) Despite several shortcomings of the FOF halo finder, e.g.,
halo-bridging (see, e.g., Gelb \& Bertschinger 1994, Summers et
al. 1995) or statistical biases found by Warren et al.~(2005), the FOF
algorithm itself is well-defined and very fast.

There are two sources of possible biases in determining individual
halo masses using FOF. First, the halo may be sampled with an
insufficient number of particles (see Warren et al. 2005).  Second,
the effective slope of the halo density profile close to the virial
radius $r_{vir}$, at fixed particle number, also influences the FOF
mass. If the force resolution of the N-body code affects the profile,
this too, adds a systematic bias. Here we record the mass function for
the linking length $b=0.2$ FOF mass including only the correction of
Warren et al. (2005). In a follow-up paper~(Luki\'c et al. 2006) we
will address systematics issues in determining halo masses in detail.
 
We now discuss criteria found to be very important for
demonstrating the convergence and robustness of our results. Details
will be presented in Luki\'c et al. (2006). The first issue relates to
the initial redshift of the simulation. Two conditions are important:
(i) the simulation must begin sufficiently early that the initial
Zel'dovich displacement is a small enough fraction of the mean
interparticle separation $\Delta_p$; on average a particle should not
move more than $\sim\Delta_p/3$; (ii) the highest redshift where the
mass function is to be evaluated must be sufficiently removed from the
redshift of first-crossing $z_{cross}$ where particles have the first
chance to form halos. The stringency of these criteria is such that
the small boxes require very high starting redshifts, e.g., the
4$h^{-1}$Mpc box had an initial redshift $z_{in}=500$. This is a much
earlier starting redshift than those used in previous simulations; the
conventional requirement that all modes in the box be linear at the
initial redshift proves to be much weaker, and therefore inadequate,
as a convergence criterion.

\begin{figure}
\begin{center}
  \includegraphics[width=65mm]{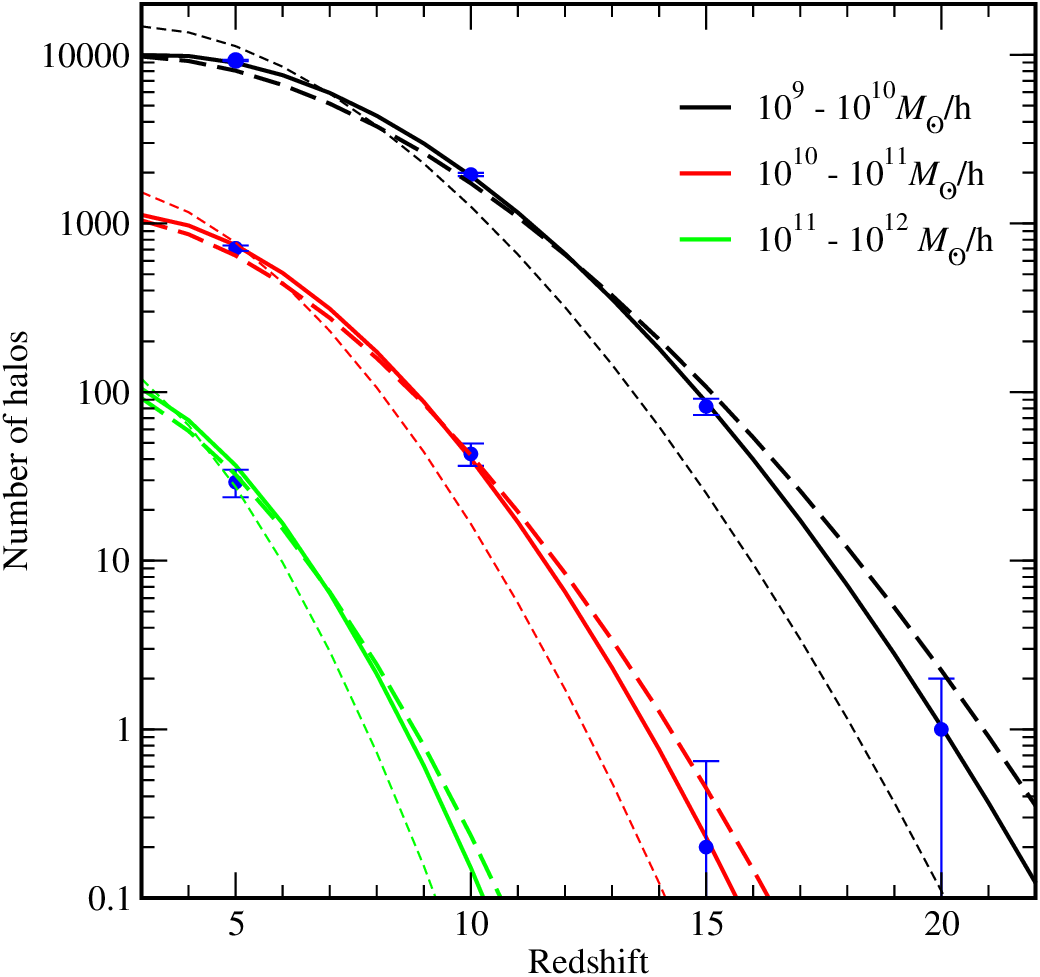}
\caption{Halo growth function for three mass bins for 
  the 16$h^{-1}$Mpc box. The Warren (solid), ST (long-dashed), and PS
  (short-dashed) fits are compared to simulation data with Poisson
  error bars. Note the quality of the agreement with the Warren fit.}
\label{plotone}
\end{center}
\end{figure}

A simple test of how well the simulations track the mass function
formulae is to follow the number of halos in a specified mass bin at a
given redshift. For this purpose we convert the mass function fit into
a function of $z$, defining the halo growth function as shown in
Figure~\ref{plotone}. The evolution of three mass bins is shown
as a function of $z$ along with results from the $16 h^{-1}$Mpc boxes.
The halo growth function is particularly valuable for determining when
the halos at a certain mass  should first form. This is a good
test for problems in simulations aiming to capture halos
with a given minimum mass at some redshift. An example of this is
insufficient force resolution in the base-grids of
adaptive-mesh-refinement (AMR) codes.

Once the number of particles for a simulation and a desired mass for
the smallest halo are decided, the required box size is fixed. The
force resolution needed to resolve the smallest halos has then to be
determined. Our aim here is not to precisely measure the halo profile
but simply to be certain that the total halo mass is correct. As shown
in Heitmann et al. (2005) the halo mass is a relatively robust
quantity and a simple estimate of the force resolution is all that is
needed. The force resolution must be small compared to the comoving
halo virial radius $r_{\Delta}$ (with the overdensity relative to the
critical density, $\Delta\sim 200$) at all redshifts. The resulting
inequality can be stated in the form
\begin{equation}
\frac{\delta_f}{\Delta_p} <
0.62\left(\frac{n_h\Omega(z)}{\Delta}\right)^{1/3}, 
\label{fres}
\end{equation}
where $\delta_f$ is the force resolution and $n_h$ is the number of
particles per halo. In the simulations performed here we use a ratio
of one particle per 64 grid cells, which allows halos with roughly 50
particles to be captured. It has been shown in Heitmann et al. (2005)
that this ratio does not cause collisional effects and leads to
consistent results in comparison with other codes. Mass function
convergence tests with different force resolutions are nicely
consistent with the above estimate as shown in Luki\'c et al. (2006);
time-step criteria and convergence tests are also described there.

\begin{figure}[t]
\begin{center}
  \includegraphics[width=80mm]{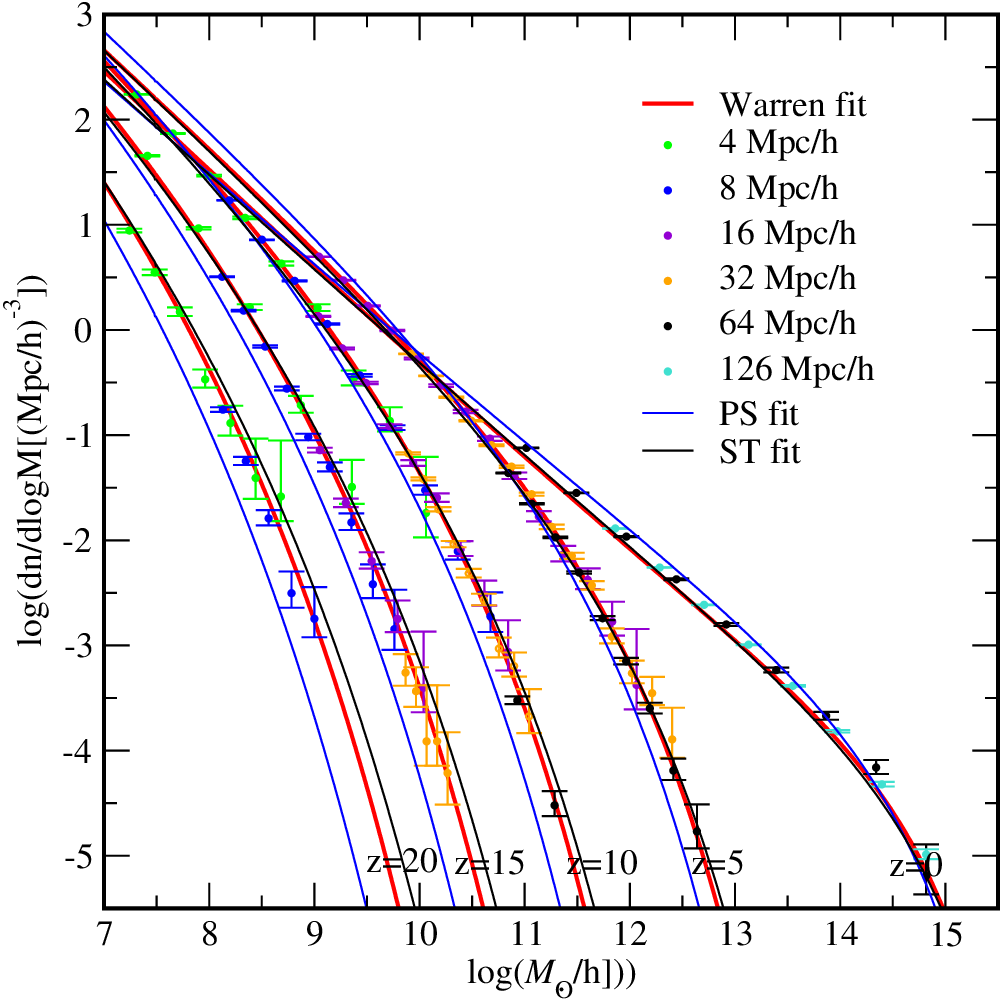}
\end{center}
\caption{The mass function at 5 different redshifts with Poisson error
bars. The red line is the Warren fit, blue is Press-Schechter, and black
is Sheth-Tormen.} 
\label{massfplot}
\end{figure}

The large set of simulations we have carried out allows us to study
the mass function at redshifts between $z=20$ and $z=0$. The main
results are shown in Figure~\ref{massfplot}, where the simulation data
for the mass function are shown along with the Warren, PS, and ST fits
at different redshifts. At all redshifts the Warren fit has the best
agreement with the simulations with a scatter of approximately $20\%$
and is a numerically significant improvement over ST. Such a close
match is quite gratifying given the overall dynamic range of the
investigation. The PS fit, over the mass range considered, is a poor
fit at $z\ge 10$, deviating by more than a factor of two from the
numerical results.

\section{Conclusions and Discussion}
\label{conclusion}
In this paper we have studied the evolution of the mass function
starting at a redshift of $z=20$ and covering a halo mass range of
$10^7$ to $10^{15}h^{-1}M_\odot$. Our results incorporate new
halo-based N-body error control criteria that are described in more
detail in Luki\'c et al. (2006). We find that the Press-Schechter
mass function deviates significantly from our results. More recent
mass function fits are in better agreement; in particular, the
recently introduced fitting function of Warren et al. (2005) agrees at
the $20\%$ level over the entire redshift range.

The precise agreement of the numerically obtained halo growth function
as well as the evolution of the mass function with the (evolved $z=0$)
Warren fit demonstrates the remarkable result that the evolution of the
mass function is completely controlled by the linear growth of the
variance of the linear density field.

In order to find a mass function fit relevant to observations, several
hurdles remain to be overcome, including reaching agreement on an
appropriate definition of halo mass (White 2001) and improving the
precision and accuracy of N-body codes beyond the current state of the
art (O'Shea et al. 2005; Heitmann et al.  2005). Depending on the
level of precision required, as White (2002) points out, ``it may not
be sufficient to use a simple parametrized form'' in constraining
cosmological parameters with the mass function.

The error control criteria developed in this work have a natural
application in high-resolution AMR simulations in the setting of
refinement and error control criteria. Work in this direction is in
progress. 

\acknowledgements

We thank Kev Abazajian, Dan Holz, Lam Hui, Gerard Jungman, Savvas
Koushiappas, Adam Lidz, Sergei Shandarin, Ravi Sheth, and Mike Warren
for useful discussions. The authors acknowledge support from IGPP,
LANL. S.H. and K.H. acknowledge support from the DOE via the LDRD
program at LANL. P.M.R. and Z.L. acknowledge UIUC, NCSA, and a
DOE/NNSA PECASE award (LLNL B532720). S.H., K.H., and P.M.R.
acknowledge the hospitality of the Aspen Center for Physics where part
of this work was carried out. We especially acknowledge supercomputing
support under the LANL Institutional Computing Initiative.

\end{document}